\documentclass[preprint,12pt]{elsarticle}

\usepackage{amssymb}
\journal{Physics Letters A}

\begin{document}

\begin{frontmatter}

%% Title, authors and addresses

%% use the tnoteref command within \title for footnotes;
%% use the tnotetext command for theassociated footnote;
%% use the fnref command within \author or \address for footnotes;
%% use the fntext command for theassociated footnote;
%% use the corref command within \author for corresponding author footnotes;
%% use the cortext command for theassociated footnote;
%% use the ead command for the email address,
%% and the form \ead[url] for the home page:
%% \title{Title\tnoteref{label1}}
%% \tnotetext[label1]{}
%% \author{Name\corref{cor1}\fnref{label2}}
%% \ead{email address}
%% \ead[url]{home page}
%% \fntext[label2]{}
%% \cortext[cor1]{}
%% \affiliation{organization={},
%%             addressline={},
%%             city={},
%%             postcode={},
%%             state={},
%%             country={}}
%% \fntext[label3]{}

\title{Tunneling time from spin fluctuations in Larmor clock}

%% use optional labels to link authors explicitly to addresses:
%% \author[label1,label2]{}
%% \affiliation[label1]{organization={},
%%             addressline={},
%%             city={},
%%             postcode={},
%%             state={},
%%             country={}}
%%
%% \affiliation[label2]{organization={},
%%             addressline={},
%%             city={},
%%             postcode={},
%%             state={},
%%             country={}}

\author[inst1]{Durmu{\c s} Demir}

\affiliation[inst1]{
organization={Sabanc{\i} University},
            addressline={Faculty of Engineering and Natural Sciences}, 
            city={Tuzla},
            postcode={34956}, 
            state={Istanbul},
           country={Turkey}}

\begin{abstract}
Tunneling time, time needed for a quantum particle to tunnel through a potential energy barrier, can be measured by a duration marker.  One such marker is  spin reorientation due to Larmor precession. With a weak magnetic field in $z$ direction, the Larmor clock reads two times, $\tau_y$ and $\tau_z$, for a potential energy barrier along the $y$ axis. The problem is to determine the actual tunneling time (ATT). B{\"u}ttiker defines $\sqrt{\tau_y^2 + \tau_z^2}$ to be the ATT. Steinberg and others, on the other hand, identify $\tau_y$ with the ATT. The B{\"u}ttiker and Steinberg times are based on average spin components but in non-commuting spin system average of one component requires the other two to fluctuate. In the present work, we study the effects of spin fluctuations and show that the ATT can well be $\tau_y + \frac{\tau_z^2}{\tau_y}$. We analyze the ATT candidates and reveal that the fluctuation-based ATT acts as a transmission time in all of the low-barrier, high-barrier, thick-barrier and classical dynamics limits.  We extract this new ATT using the most recent experimental data by the Steinberg group. The new ATT qualifies as a viable tunneling time formula. 
\end{abstract}

%%Graphical abstract
%\begin{graphicalabstract}
%\includegraphics{grabs}
%\end{graphicalabstract}

%%Research highlights
%\begin{highlights}
%\item Research highlight 1
%\item Research highlight 2
%\end{highlights}

\begin{keyword}
%% keywords here, in the form: keyword \sep keyword
Larmor time \sep Spin uncertainty \sep Tunneling time
%% PACS codes here, in the form: \PACS code \sep code
%\PACS 0000 \sep 1111
%% MSC codes here, in the form: \MSC code \sep code
%% or \MSC[2008] code \sep code (2000 is the default)
%\MSC 0000 \sep 1111
\end{keyword}

\end{frontmatter}

%% \linenumbers

%% main text
\section{Introduction}
Tunneling is the transmission of quantum particles through potential energy barriers which exceed their total energy \cite{gamow,condon}. It is a pure quantum effect underlying various natural phenomena and technological devices. Its duration, the tunneling time, has been formulated by utilizing various observables with various methods \cite{time1,review00,review01,review02}. The reason for the existence of various tunneling time formulas is that the role of time in quantum theory is a controversial issue. There is the notion of observable time (flight time) hampered by Pauli's famous theorem \cite{pauli,review03,busch}. There is also the notion of external time related to surroundings of the quantum system \cite{busch,hilgevoord}. There is again the notion of intrinsic time that can be marked by an observable intrinsic to the quantum system. It is this intrinsic time which will be studied in the present work. By definition, march of the intrinsic time can be parametrized by using an appropriate marker namely an observable representative of the quantum system. The marker could be phase shift \cite{wigner}, spin precession \cite{larmor1,larmor2}, entropy production \cite{demir-guner} or any other change pertinent to the tunneling particle. The spin precession across feebly-magnetized potential barriers, for instance, meters the tunneling time without altering energetics of the particle. This marker, the Larmor clock \cite{buttiker1,buttiker2,buttiker3}, reads two times, $\tau_y$ and $\tau_z$,  with a feeble magnetic field in the $z$ direction and a potential energy barrier along the $y$ axis. These times have recently been measured by Steinberg group using the rubidium atoms \cite{larmor-rubidium}. There is, in general, no telling what combination of $\tau_y$ and $\tau_z$ forms the actual tunneling time (ATT) \cite{buttiker1,reaction2,reaction3,review03}.  B{\"u}ttiker defines $(\tau_y^2 + \tau_z^2)^{1/2}$ to be the ATT. Steinberg and others, on the other hand, identify $\tau_y$ with the ATT after attributing $\tau_z$ to measurement back-action \cite{reaction1,reaction2,larmor-rubidium}. The key point about $\tau_y$ and $\tau_z$ is that they are based on average spin components, and do not therefore involve possible contributions from spin fluctuations.

The Larmor precession times $\tau_y$ and $\tau_z$ are set by the average spin components. They do not carry information on how the transmitted spins are distributed about their averages. Their distributions are important because different spin orientations imply different precession times. In this sense, variances of the spins normalized to their average values (namely the Fano factors \cite{fano,fano2} for spin components) act as a measure of how dispersed the spin components get with respect to their average values in course of the tunneling. Fano factor for a given spin component is equal to $\hbar/2$ if spins are Poisson-distributed, larger if spins are clustered, and smaller if spins are uniformly-distributed \cite{stat-paper}. Needless to say, each spin distribution leads to a precession time distribution of its own, and, in this sense, the Fano factor is expected to lead to a factual determination of the ATT. 

In the present work, we study the effects of spin fluctuations on Larmor clock. We  find that spin uncertainty, whose relevance is hinted by the spin component in the direction of the magnetic field,  gives rise to a new tunneling time formula. This fluctuation-induced ATT applies directly to experimental data \cite{larmor-rubidium}, and differs from the B{\"u}ttiker and Steinberg ATT definitions in terms of the transmission speeds involved.  In Sec. 2 below we derive the ATT from spin fluctuations, and contrast it with the B{\"u}ttiker and Steinberg definitions in asymptotic limits. In Sec. 3 we analyze the fluctuation-induced ATT we constructed using the available experimental data \cite{larmor-rubidium}, and discuss its physics implications considering the relevant limit values.  In Sec. 4 we conclude with a discussion of the fluctuation-induced ATT  in regard to its potential applications like the quantum annealing \cite{anneal5,chip}.  

\section{Actual Tunneling Time from Spin Fluctuations}

Let us consider an ensemble of identical particles each having mass $m$, magnetic moment $\mu$ and gyromagnetic ratio $g$. The ensemble is endowed with  average energy $E$ and average spin components
\begin{eqnarray}
\langle S_x\rangle = \hbar/2\,,\; \langle S_y \rangle= 0\,,\; \langle S_z\rangle= 0
\label{spinsi}
\end{eqnarray}
as sharply-peaked distributions with small variance-to-mean ratios. This ensemble moves along the $y$-axis with average momentum $\sqrt{2 m E}$ to scatter at a potential energy profile  $V(y)$, which confines a uniform magnetic field $B$ in the $z$-direction. The particles get eventually either reflected or transmitted. Under the effect of the magnetic field, 
 the transmitted particles attain the average spin components \cite{buttiker1,buttiker2,buttiker3} 
\begin{eqnarray}
\langle S_x \rangle = \hbar/2\,,\; \langle S_y \rangle = (\hbar/2) \omega_L \tau_y\,,\; \langle S_z \rangle = (\hbar/2) \omega_L \tau_z 
\label{spinst}
\end{eqnarray}
up to ${\cal{O}}(\omega_L^2)$ accuracy in the Larmor frequency $\hbar \omega_L = g\mu B$ 
in a feeble magnetic field $(\hbar \omega_L \ll E)$. This transmitted spin differs from the incident spin in (\ref{spinsi}) by its nonzero $y$- and $z$-components, and gives rise therefore to two distinct precession times, $\tau_y$ and $\tau_z$. The time $\tau_y$ is a direct consequence of the magnetic field because it remains nonzero even when the potential is absent.  The time $\tau_z$, on the other hand, results from the potential in that it vanishes in the limit of vanishing potential. There is, nevertheless, no telling what combination of  $\tau_y$ and $\tau_z$ is the ATT. In fact, in his seminal work \cite{buttiker1}, B{\"u}ttiker defines the ATT to be 
\begin{eqnarray}
({\rm ATT})_{B} = \sqrt{\tau_y^2 + \tau_z^2}
\label{att-b}
\end{eqnarray}
as the quadrature sum of the $\tau_y$ and $\tau_z$. In the weak measurement formalism \cite{reaction1}, on the other hand, Steinberg takes the ATT to be 
\begin{eqnarray}
({\rm ATT})_{S} = \tau_y
\label{att-s}
\end{eqnarray}
by attributing $\tau_z$ to measurement back-action \cite{reaction2,reaction3}. There are thus two proposed formulas for the duration of tunneling through a given barrier. Below, we construct a third formula based on spin fluctuations.

The definitions of $\tau_y$ and $\tau_z$ in (\ref{spinst}) make it clear that both $({\rm ATT})_{B}$ and $({\rm ATT})_{S}$ are based on the average spin components $\langle S_{y,z}\rangle$. It is, however,  not possible to measure any one of the  spin components without disrupting the other two. This follows from  non-commuting nature of the spin components. It is in this sense that the spin uncertainty comes to the fore, and puts $\langle S_z\rangle$ at the center stage in agreement with its role as the measurement back-action in the weak measurement formalism \cite{reaction2,reaction3}. In fact,  $\langle S_z\rangle$ is a sensitive probe of the potential and, at the same time,  underpins the uncertainly relation
\begin{eqnarray}
(\Delta S_x)^2 (\Delta S_y)^2 \geq \frac{\hbar^2}{4} \langle S_z \rangle^2
\label{uncertainty}
\end{eqnarray}
according to which the $S_x$ and $S_y$ variances ($i=x,y$)
\begin{eqnarray}
(\Delta S_{i})^2 = \langle (S_{i}-\langle S_{i}\rangle)^2 \rangle = \langle S_{i}^2 \rangle - \langle S_{i}\rangle^2  
\label{variances}
\end{eqnarray}
are related to each other seesawically, with a pivot at $\langle S_z \rangle$. 
In view of these variances, the ratio  $(\Delta S_{x,y})^2/\langle S_{x,y}\rangle$, which is known as the Fano factor \cite{fano,fano2}, acts a measure of how dispersed the spin components get with respect to their average values in course of the transmission (tunneling). It is equal to $\hbar/2$ if spins are Poisson-distributed, larger if spins are clustered, and smaller if spins are uniformly-distributed \cite{stat-paper}. 

The uncertainty product $(\Delta S_x)^2 (\Delta S_y)^2$ in (\ref{uncertainty}) is expected to have a strong correlation with the potential barrier. This follows from sensitivity of $\langle S_z \rangle$ to the  potential profile, and  comes to mean that the time delay caused by the potential, the sought-for ATT itself, must actually be encoded in the product $(\Delta S_x)^2 (\Delta S_y)^2$ via $\langle S_z \rangle$. In this regard, we first reorganize $(\Delta S_x)^2 (\Delta S_y)^2$ as normalized Fano factors. Next, we observe that the sought-for ATT (transmission time), irrespective of how it is formulated, must be measured in terms of the inverse Larmor frequency $\omega_L^{-1}$ because it is the only time scale in the problem. In view of these features,  spin fluctuations can be conjectured to give rise to a new ATT 
\begin{eqnarray}
({\rm ATT})_{F}  = \omega_L^{-1}\times \frac{(\Delta S_x)^2}{\frac{\hbar}{2} \langle S_x \rangle} \times \frac{(\Delta S_y)^2}{\frac{\hbar}{2} \langle S_y \rangle} 
\label{dis-dur}
\end{eqnarray}
in which the subscript ``F" is a reminder that this new ATT is based on spin ``fluctuations". Up to ${\mathcal{O}}(\omega_L^2)$ accuracy under a feeble magnetic field $(\hbar \omega_L \ll E)$, the spin variances in  (\ref{dis-dur}) take the forms \cite{buttiker1}
\begin{eqnarray}
(\Delta S_x)^2 &=& \frac{\hbar^2}{4} \omega_L^2 (\tau_y^2 + \tau_z^2)\\
(\Delta S_y)^2 &=& \frac{\hbar^2}{4} \left(1- \omega_L^2 \tau_y^2 \right)
\label{new-variances}
\end{eqnarray}
after using the spin averages in (\ref{spinst}) in the formula (\ref{variances}) for spin variances by taking into account the fact that $\langle S_i^2 \rangle = {\hbar^2}/{4}$ (for each $i=x,y,z$) by the property of the Pauli spin matrices.  Now, using these spin variances in (\ref{dis-dur}) along with the spin averages in (\ref{spinst}) one is led to the  fluctuation-induced transmission time formula 
\begin{eqnarray}
({\rm ATT})_{F} = \tau_y  + \frac{\tau_z^2}{\tau_y}
\label{spin-time}
\end{eqnarray}
under a feeble magnetic field $(\hbar \omega_L \ll E)$ for which ${\cal{O}}(\omega_L^2)$ terms in (\ref{dis-dur}) can all be dropped, and average spin components in (\ref{spinst}) can be measured without disrupting the state of the particle \cite{reaction1,reaction2,reaction3}. (It is necessary to suppress ${\mathcal{O}}(\omega_L^2)$ contributions as they enhance also orbital effects.) It differs from the earlier transmission time  definitions \cite{buttiker1,reaction1,reaction2} not only by its functional form but also by its base rock of spin uncertainty. It has general validity in that it is independent of what shape \cite{buttiker2,buttiker3} the potential has, how wide or high the tunneling region $(V(y) > E)$ is, and if the potential is time-dependent \cite{elci} or not \cite{buttiker2,buttiker3}.  It can be extracted directly from $\tau_y$ and $\tau_z$ data irrespective of if these two precession times are measured by experiment or modeled by theory. 

\begin{table}
\caption{\label{table1}The three ATT candidates in the low-barrier, high-barrier, thick-barrier and classical dynamics limits.}
\begin{tabular}{|l|l|l|l|l|l|l|}
\hline
 & $\tau_y$ & $\tau_z$ & $({\rm ATT})_{B}$ & $({\rm ATT})_{S}$ & $({\rm ATT})_{F}$  \\
\hline
\hline
\begin{tabular}{l}
low-barrier: $V_0\ll E$\\
(fixed $E$)  
\end{tabular} & $\tau_c(0,E)$ & $0$ & $\tau_c(0,E)$& $\tau_c(0,E)$ & $\tau_c(0,E)$
\\
\hline
\begin{tabular}{l}
high-barrier: $E\ll V_0$\\
(fixed $V_0$)  
\end{tabular} & $0$ & $\tau_c(V_0,0)$ & $\tau_c(V_0,0)$& $0$ & $\infty$
\\
\hline
\begin{tabular}{l}
thick-barrier: $L^2\gg \frac{\hbar}{m} \tau_c(V_0,E)$\\
(fixed $V_0,E$)  
\end{tabular} & $\frac{\hbar}{V_0}\frac{\tau_c(V_0,E)}{\tau_c(0,E)}$ & $\infty$ & $\infty$& $\frac{\hbar}{V_0}\frac{\tau_c(V_0,E)}{\tau_c(0,E)}$ & $\infty$
\\
\hline
\begin{tabular}{l}
classical dynamics: $\hbar\rightarrow 0$\\
(fixed $V_0,E,L$)  
\end{tabular} & $0$ & $\tau_c(V_0,E)$ & $\tau_c(V_0,E)$& $0$ & $\infty$
\\
\hline
\end{tabular}
\end{table}
There are three ATT candidates $({\rm ATT})_{B}$, $({\rm ATT})_{S}$ and $({\rm ATT})_{F}$. The question of which ATT is realized in nature will eventually be answered by experiment. In the absence of experimental data, all one can do is to contrast these ATTs in  physically discriminate configurations. To this end, a rectangular potential barrier of height $V_0$ and width $L$ proves particularly eligible. Indeed, this configuration admits a complete analytic illustration and, as a result,  one finds that \cite{buttiker1,buttiker2}
\begin{eqnarray}
\tau_y =\frac{\hbar}{2} \left(\frac{E}{V_0-E}\right)^{1/2} 
\frac{2 \lambda (V_0 - 2 E)+ V_0 \sinh 2 \lambda}{4 E (V_0-E)  + V_0^2 
\sinh^2\lambda}
\label{tauy}
\end{eqnarray}
and
\begin{eqnarray}
\tau_z = \frac{\hbar}{4} \frac{V_0}{V_0-E} \frac{2 (V_0 - 2 E) 
\sinh^2\lambda+ V_0 \lambda \sinh 2 \lambda}{4 E (V_0-E)  + V_0^2 \sinh^2\lambda}
\label{tauz}
\end{eqnarray} 
with the barrier opacity  $\hbar \lambda = L (2m (V_0-E))^{1/2}$. The ATT candidates  $({\rm ATT})_{B}$, $({\rm ATT})_{S}$ and $({\rm ATT})_{F}$ derive from $\tau_y$ and $\tau_z$ via their definitions in (\ref{att-b}), (\ref{att-s}) and (\ref{spin-time}). Their physics implications are best revealed by contrasting them in their asymptotic forms. Indeed, their forms in the low-barrier, high-barrier, thick-barrier and classical dynamics limits can shed light on their capabilities as transmission times. In this regard, they are compared in Table \ref{table1} for each of these limits. In the low-barrier limit ($E/V_0 \gg 1$), they all agree and give a unique ATT 
\begin{eqnarray}
\tau_c(V_0,E)=\frac{m L}{\sqrt{2 m|V_0-E|}}
\label{time-class}
%\quad \, {\rm and}\;\quad \tau_c(0,E)=\frac{m L}{\sqrt{2 m E}}
\end{eqnarray}
designating the time it takes for a classical particle to traverse a distance $L$ with effective energy  $|V_0-E|$ \cite{buttiker1,reaction2,buttiker-landauer}. This agreement of the three ATT candidates, as given in Table \ref{table1},  stems from vanishing of $\tau_z$ for low-potentials. It is in this sense that $\tau_z$ is a sensitive probe of the potential. It qualifies to be a viable measure of the march of time in the presence of the potential barrier. 

The potential barrier becomes opaque ($\lambda \gg 1$) in the limit of thick and high barriers. It becomes opaque in the limit also of the  a classical dynamics. In this opaque limit, the three ATT candidates take the asymptotic forms  
\begin{eqnarray}
\lim_{\lambda\rightarrow\infty} ({\rm ATT})_{B} &=& \sqrt{\frac{\hbar^2}{V^2_0} \frac{\tau^2_c(V_0,E)}{\tau^2_c(0,E)} + \tau^2_c(V_0,E)},\nonumber\\
\lim_{\lambda\rightarrow\infty} ({\rm ATT})_{S} &=& \frac{\hbar}{V_0} \frac{\tau_c(V_0,E)}{\tau_c(0,E)} \,,\label{limits}\\
\lim_{\lambda\rightarrow\infty} ({\rm ATT})_{F} &=& \frac{\hbar}{V_0} \frac{\tau_c(V_0,E)}{\tau_c(0,E)} + \frac{V_0 \tau_c(V_0,E)\tau_c(0,E)}{\hbar},\nonumber
\end{eqnarray}
in which $\tau_c(0,E)\equiv \tau_c(V_0=0,E)$ where $\tau_c(V_0,E)$ is defined in (\ref{time-class}). These high-opacity limits encompass each of the classical, thick-barrier and high-barrier limits, as elaborated below:
\begin{enumerate}
\item In the high-barrier limit ($E/V_0 \ll 1$),  $({\rm ATT})_{B}$ and $({\rm ATT})_{S}$, as given in Table \ref{table1}, exhibit  non-quantum behavior because they both remain finite for tunneling through the barrier. This is not expected in quantum mechanics  since wavefunction is diminished in the barrier region, and this diminishing suppresses the transmission exponentially to give cause to a long transmission (tunneling) time \cite{buttiker1,buttiker2,buttiker3}. In contrast to $({\rm ATT})_{B}$ and $({\rm ATT})_{S}$, the fluctuation-based tunneling time $({\rm ATT})_{F}$ becomes infinitely long and this is in perfect agreement with the quantum expectations.

\item In the thick-barrier limit ($L^2\gg \frac{\hbar}{m} \tau_c(V_0,E)$), as given in Table \ref{table1},  $({\rm ATT})_{S}$ becomes width-independent, and $({\rm ATT})_{B}$ and  $({\rm ATT})_{F}$ both tend to infinity. It is clear that $({\rm ATT})_{S}$ exhibits again non-quantum behavior. The longevity of $({\rm ATT})_{B}$ and $({\rm ATT})_{F}$, on the other hand, is precisely what is expected of transmission (tunneling) time in view again of the suppression of the transmission probability under the barrier \cite{buttiker1,buttiker2,buttiker3}. It is possible to infer that $({\rm ATT})_{S}$ can pertain to group velocity since group velocity can be superluminal. The other two, $({\rm ATT})_{B}$ and  $({\rm ATT})_{F}$, on the other hand, could result from wave-front velocity (the subluminal speed with which waves carry information) \cite{velocity,forefront1,forefront2}. 

\item In the classical particle limit ($\hbar \rightarrow 0$), as given in Table \ref{table1}, $({\rm ATT})_{B}$ and $({\rm ATT})_{S}$ exhibit non-classical behavior as they both remain finite. This is not expected in classical mechanics in which tunneling through barriers is simply impossible. The fluctuation-induced tunneling time $({\rm ATT})_{F}$ becomes infinitely long. This longevity is precisely what is expected of transmission (tunneling) time for a classical particle  \cite{gamow,review00,review03}. 
\end{enumerate}
These asymptotic behaviors lead to an unambiguous conclusion which is that the $({\rm ATT})_{F}$ possesses all the features expected of a transmission (tunneling) time. On physical grounds, therefore, $({\rm ATT})_{F}$ is expected to be a serious ATT candidate, possibly pertaining to wave-front speed \cite{velocity,forefront1,forefront2}. 

\section{Extracting Actual Tunneling Time from Experimental Data}

\begin{figure}[h!]
\includegraphics[scale=0.42]{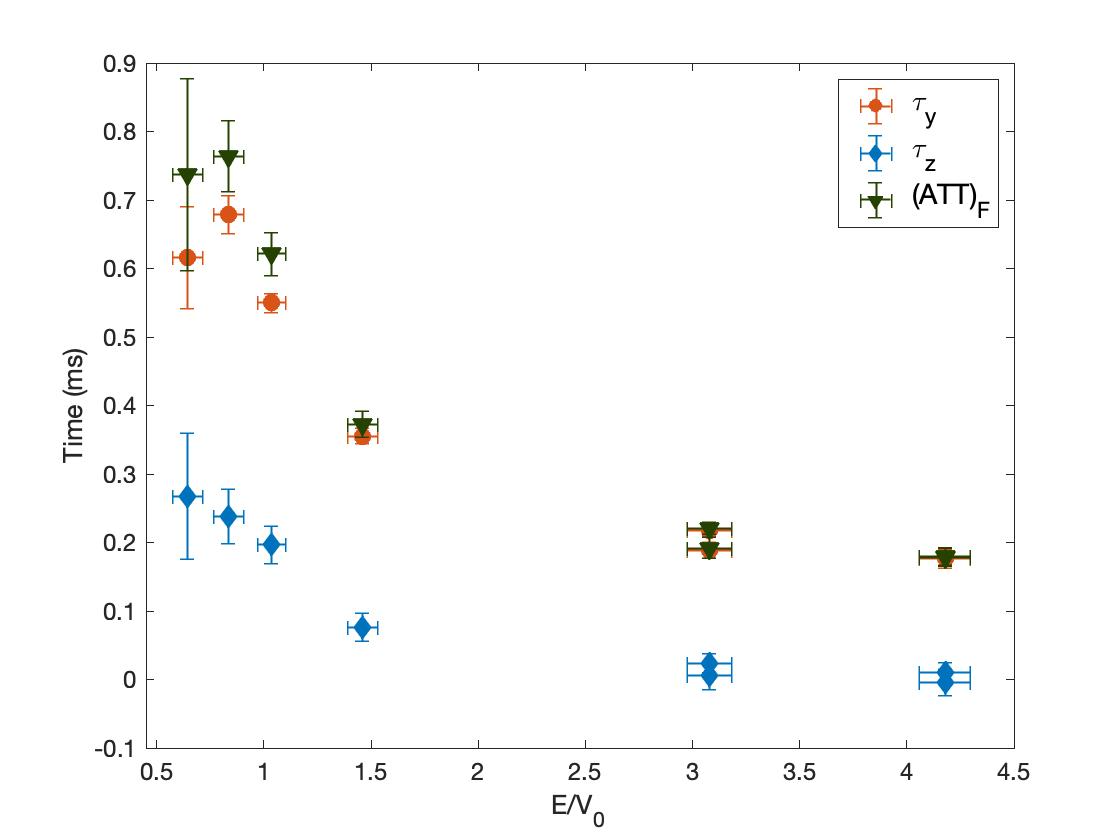}    
\caption{The $({\rm ATT})_F$ (green/triangle) as derived via (\ref{spin-time}) from the experimental data \cite{larmor-rubidium} on precession times $\tau_y$ (red/bullet) and $\tau_z$ (blue/diamond) of the  $^{87}$Rb isotopes transmitted with average energy $E$ through a Gaussian barrier of peak value $V_0=135\ {\rm nK}$ and width  $L_0=1.3\ {\rm \mu m}$.  There is an indistinct trend that, in the deep tunneling regime $(E\ll V_0)$,  $\tau_y$ $(\tau_z)$ can decrease (increase) and, as a result, the $({\rm ATT})_F$ can take large values.}
\label{figure-time-exp}
\end{figure}

In this section, we shall extract $({\rm ATT})_{F}$ from the available experimental data, and discuss its asymptotics. Experiments with inert gases have already given strong evidence for finite tunneling time 
\cite{evidence1,evidence2,evidence3}. These atomic evidences are hardly eligible for model building since it is not possible to know the start time of the electron tunneling \cite{demir-guner}. The experiments with cold atoms have turned these evidences into accurate measurements. Indeed, $\tau_y$ and $\tau_z$ have recently been measured by Steinberg group \cite{larmor-rubidium} on an ensemble of some 8,000 Bose-condensed $^{87}$Rb isotopes. In the experiment, the ensemble is staged \cite{larmor-rubidium0,larmor-rubidium1,larmor-rubidium2}  to scatter at an optical Gaussian barrier of peak value $V_0=135\ {\rm nK}$ and width  $L_0=1.3\ {\rm \mu m}$, and spin components of the transmitted isotopes are measured via their absorption spectra. The experimental data agrees with basic theoretical predictions (Table \ref{table1}) in that, for low potentials $(E/V_0 \rightarrow \infty)$, $\tau_z$ vanishes 
$(\tau_z \rightarrow 0)$ but $\tau_y$ remains finite
$(\tau_y \rightarrow m L_0/\sqrt{2 m E})$. The data does not extend to slower isotopes but still it exhibits an indistinct trend that, for high potentials $(E/V_0 \rightarrow 0)$, $\tau_y$ vanishes $(\tau_y\rightarrow 0)$ but $\tau_z$ remains finite. The actual transmission duration of the  $^{87}$Rb atoms, modeled by the $({\rm ATT})_F$ in (\ref{spin-time}) as a fluctuation-induced tunneling time, lies above both $\tau_y$ and $\tau_z$ on the average, and has error bars of similar size as theirs (Fig. \ref{figure-time-exp}).  The $({\rm ATT})_F$ approaches to $\tau_y$ at low potentials $(E/V_0\rightarrow \infty)$ and shows a slight trend that it may blow up at high potentials $(E/V_0\rightarrow 0)$. 

\begin{figure}[h!]
\includegraphics[scale=0.62]{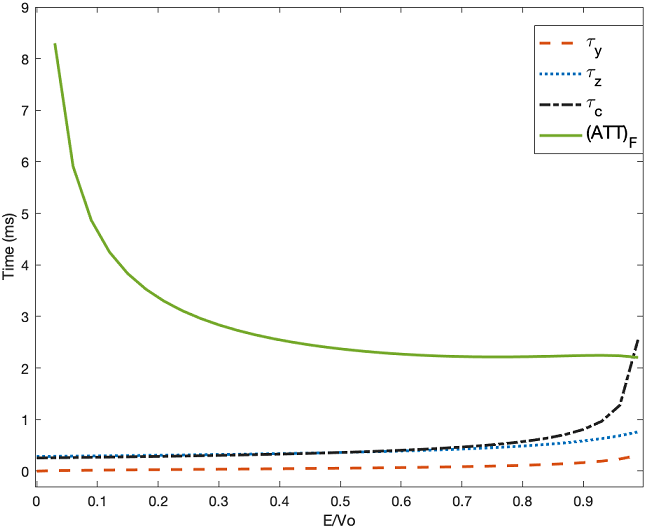}
\caption{The $({\rm ATT})_F$ (green/full curve) as derived via (\ref{spin-time}) with $\tau_y$ in (\ref{tauy}) (red/dashed curve) and  $\tau_z$ in (\ref{tauz}) (blue/dotted curve) evaluated for  $^{87}$Rb isotopes transmitted with average energy $E$ through a rectangular barrier of height $V_0=135\ {\rm nK}$ and width  $L_0=1.3\ {\rm \mu m}$. Also depicted is the classical time $\tau_c$ in (\ref{time-class}) (black/dot-dashed curve). For high barriers $(E/V_0\rightarrow 0)$, $\tau_y$ vanishes, $\tau_z$ decreases but remains finite (Table \ref{table1}) and, as a result, the $({\rm ATT})_F$ blows up, showing that it takes forever to tunnel though a sky-high barrier. In contrast to $({\rm ATT})_F$, the classical time $\tau_c$ blows up as $E/V_0 \rightarrow 1$.}
\label{figure-time-th-EbV} 
\end{figure}

\begin{figure}[h!]
\includegraphics[scale=0.8]{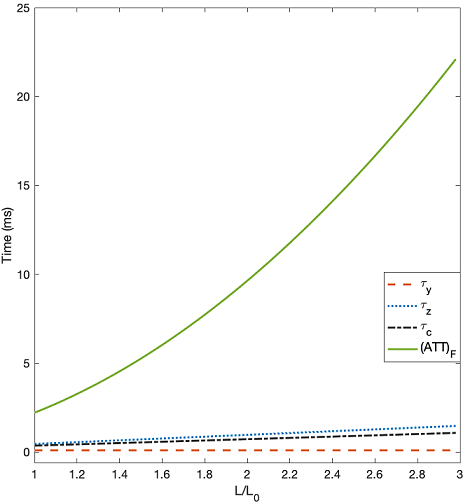}
\caption{The $({\rm ATT})_F$ (green/full curve) as derived via (\ref{spin-time}) with $\tau_y$ in (\ref{tauy}) (red/dashed curve) and  $\tau_z$ in (\ref{tauz}) (blue/dotted curve) evaluated for $^{87}$Rb isotopes transmitted with average energy $E=V_0/2$ through a rectangular barrier of height $V_0=135\ {\rm nK}$ and width $L$ (in units of $L_0=1.3\ {\rm \mu m}$). Also depicted is the classical time $\tau_c$ in (\ref{time-class}) (black/dot-dashed curve).
For wide barriers $(L/L_0\rightarrow \infty)$, $\tau_y$ becomes width-independent, $\tau_z$ increases slightly (Table \ref{table1}), $\tau_c$ grows linearly and yet the $({\rm ATT})_F$ grows exponentially, showing that it takes forever to tunnel though a worlds-wide barrier.}
\label{figure-time-th-LbL0} 
\end{figure}

The way the $({\rm ATT})_F$, $\tau_y$ and $\tau_z$ approach to the high and wide barrier limits in (\ref{limits}) are important not only for testing the models  but also for guiding future experiments (Table \ref{table1}). Indeed, transcribing the lab specs \cite{larmor-rubidium}  of  $V_0=135\ {\rm nK}$ and $L_0=1.3\ {\rm \mu m}$ to the rectangular potential under concern, it is found that the $({\rm ATT})_F$ gradually increases and eventually diverges as $E/V_0 \rightarrow 0$, and ensures this way that high barriers are hard to tunnel through (Fig. \ref{figure-time-th-EbV}). The experimental $({\rm ATT})_F$ (Fig. \ref{figure-time-exp}), too, is expected to exhibit a similar divergence if future experiments probe the deep tunneling regime $(E\ll V_0)$ beyond the available data \cite{larmor-rubidium}. In contrast to the $({\rm ATT})_F$, $\tau_y$, the transmission time $({\rm ATT})_S$ \cite{reaction2,reaction3} in weak measurement formalism \cite{reaction1},  makes the unphysical prediction that tunneling must take zero-time for macroscopic bodies and high barriers. In contrast, the classical time $\tau_c$ blows up as $E/V_0 \rightarrow 1$ kinematically  (Table \ref{table1} and Fig. \ref{figure-time-th-EbV}).

The wide barrier limit (Table \ref{table1}) and the way it is approached (Fig. \ref{figure-time-th-LbL0}), yet to be probed by experiments like neutron tunneling through isotopic nanostructures \cite{neutron}, reveal that the $({\rm ATT})_F$ involves slower-than-light transmission, and is thus free of the Hartman effect \cite{review01,hartman1}. The classical time $\tau_c$ grows linearly with the barrier width. In contrast, however, the precession time $\tau_y$, identified  with actual transmission time in weak measurement approach \cite{reaction1,reaction2,reaction3}, implies faster-than-light transmission (Table \ref{table1}), and exhibits therefore the Hartman effect \cite{hartman1}. The resolution \cite{velocity} of this unphysical behavior is that $\tau_y$ pertains to group delay rather than wave-front delay \cite{forefront1,forefront2} and, since actual transmission time must be set by the wave-front speed \cite{forefront2} faster-than-light transmission implied by $\tau_y$ is of no physical consequence. Likewise,   faster-than-light transmission observed \cite{photon1,photon2,photon3,photon4} in evanescent electromagnetic waves refers to group delay \cite{velocity,forefront1,forefront2}. Thus, in view of its physical behavior at the extremes of the potential (Table \ref{table1}), and in view also of its longevity compared to $\tau_y$ and $\tau_z$, the fluctuation-induced time $({\rm ATT})_F$ possesses every reason to qualify as wave-front delay \cite{forefront2}. 

\section{Conclusion}
In conclusion, we have shown that spin uncertainty, whose relevance is hinted by the spin component along the magnetic field, shadows forth a new tunneling time $({\rm ATT})_F$.  This new time is  a slower-than-light combination of the two Larmor times $\tau_y$ and $\tau_z$. It remains  valid for general potential profiles, and can be extracted directly from the Larmor clock readings. It is concordant with the wavefront delay, and respects therefore fundamentals of the relativity and quantum. Needless to say, $({\rm ATT})_F$ gives the durations of  tunneling-enabled phenomena in nature, which range from nuclear fusion \cite{gamow} in physics to smell perception \cite{smell} in chemistry to DNA mutation \cite{DNA1,DNA2,DNA3,DNA4} in biology. Likewise, it sets operation speeds of all tunneling-driven processes such as quantum annealing \cite{anneal1,anneal2} -- a tunneling-enabled quantum computation mechanism \cite{anneal3,anneal4,anneal5} which has already been implemented on an 1,800-qubit chip \cite{chip} by D-Wave Systems. Quantum annealing works in search spaces having numerous local minima \cite{anneal4,anneal5}, and getting out of each minimum costs a time delay given by the $({\rm ATT})_F$. In this sense, the $({\rm ATT})_F$ pops up as an additional factor to be taken into account in assessing the performance of quantum computers. This example of quantum annealing can be extended to operation speeds of various other tunneling-driven technological devices.  

\section*{Acknowledgements}
This work is supported in part by the IPS Project B.A.CF-20-02239 at Sabanc{\i} University. The author is grateful Ram\'{o}n Ramos for sharing the experimental data, G. Demir for computational help,  and O. Sarg{\i}n for discussions. 
 
\bibliographystyle{plain}
%\bibliography{references}

\end{document}